\documentstyle[preprint,prl,aps]{revtex}
\tightenlines
\begin{document}
\draft

\preprint{Hiroshima University Preprint HUBP-09/02}

\title{A new aspects of physics of a photon gas}

\author{ Levan N. Tsintsadze}
\address{Venture Business Laboratory, Hiroshima University, Higashi-Hiroshima,
Japan}

\date{November 29, 2002}

\maketitle

\begin{abstract}

Bose-Einstein condensation (BEC) and evaporation of transverse
photons from the Bose condensate is studied in the case when the
density of plasma does not change. The generation of the
longitudinal photons (photonikos) by the transverse photons
(photons) in terms of the Cherenkov type of radiation in a uniform
plasma is demonstrated. The Bogoliubov energy spectrum is derived
for photonikos.  A new physical phenomena of the "Compton"
scattering type in nonlinear photon gas is discussed. To this end,
a new version of the Pauli equation in the wavevector
representation is derived. The formation of Bose-Einstein
condensate and evaporation of photons from the condensate is
investigated by means of the newly derived Fokker-Planck equation
for photonikos. The relevance of this work to recent discovery of
black hole X-ray jets is pointed out.

\end{abstract}

\vspace{2cm}

\pacs{ \hspace{3cm} Submitted to Physical Review A}

\section{Introduction}

An immense amount of research has been carried out on propagation
of relativistically intense electromagnetic (EM) waves into an
isotropic plasma demonstrating that the relativistic oscillatory
motion of electrons causes a whole set of interesting and salient
phenomena \cite{shu}-\cite{kiv}, relevant to the study of laser
accelerators of electrons, ions and photons, laser fusion,
nonlinear optics, etc. Some of them have already been confirmed by
experiments due to the recent progress in compact, high-power,
short pulse laser technology. In our previous paper \cite{tsin} we
have suggested a new mechanism for ultrahigh gradient electron and
ion acceleration. In addition we have shown over $10^7G$ generated
magnetic fields by the nonpotential ponderomotive force.

The above treatments were restricted to the case of monochromatic
EM waves. However, the interaction of relativistically intense
radiation with a plasma leads to several kinds of instabilities
and the initially coherent spectrum may eventually broaden, and
naturally for ultrashort pulses the initial bandwidth is
increasingly broad. Moreover, in astronomical plasmas there are a
variety of sources of radiation, and in this case we speak about
the average density of radiation from all the sources and their
spectral distribution. Hence, the natural state of the strong
radiation of EM field is with a broad spectrum. In order to study
the interaction of spectrally  broad and relativistically intense
EM waves with a plasma, it was necessary to derive a general
equation for the EM spectral intensity. Such an investigation was
reported recently \cite{ltsin98}, \cite{ntsin98}, where the
authors derived a general kinetic equation for the photons in a
plasma.

It should be emphasized that the radiation can be in two distinct
states. Namely, one is when the total number of photons is not
conserved. A good example is a black-body radiation. Another
situation is when the total number of photons is conserved. Both
these states have been studied in several aspects, mostly for the
weak radiation.

It is well known that the nature of EM waves in a vacuum is quit
different from the one in a medium. In the vacuum EM wave exists
only in motion, however the light can be stopped (wavevector
$\vec{k}=0,\ \omega=\omega(0)\neq 0$) in different mediums and
wave-guides. As was shown in Ref.\cite{ltsin96} the photons
acquire the rest mass and become one of the Bosons in plasmas and
posses all characteristics of nonzero rest mass, i.e., we may say
that the photon is the elementary particle of the optical field.

Besides, a several reviews and books have been published on theory
of Bose-Einstein condensation (BEC) in a quantum Bose liquid
\cite{lif} and in trapped gases \cite{dal},\cite{ant}. Kompaneets
\cite{kom} has shown that the establishment of equilibrium between
the photons and the electrons is possible through the Compton
effect. In his consideration, since the free electron does not
absorb and emit, but only scatters the photon, the total number of
photons is conserved. Using the kinetic equation of Kompaneets,
Zel'dovich and Levich \cite{zel} have shown that in the absence of
absorption the photons undergo BEC. Such a possibility of the BEC
occurs in the case, when the processes of change of energy and
momentum in scattering dominates over the processes involving
change of the photon number in their emission and absorption.

Very recently it was shown that exists an another new mechanism of
the creation of equilibrium state and Bose-Einstein condensation
in a nonideal dense photon gas \cite{ltsin02}. More importantly a
new effect was predicted in the same paper, namely that the
inhomogeneous dense photon gas can be found in the intermediate
state.

In the present paper, we consider the BEC and evaporation of the
transverse photons (photons) from the Bose condensate. In our
study we assume that the intensity of radiation (strong and
super-strong laser pulse, non-thermal equilibrium cosmic field
radiation, etc.) is sufficiently large, so that the photon-photon
interaction can become more likely than the photon-electron
interaction. We will show that for certain conditions the
variation of the plasma density can be neglected in comparison
with the variation of the photon density. In such case the
elementary excitations represent the longitudinal photons
(photonikos), for which we will derive the well known Bogoliubov's
energy spectrum.

The paper is organized as follows. First in Sec.II a basic
equations, describing the relativistic photon-plasma interactions,
is presented. The problem of stability of the photon flow is
discussed in Sec.III. The derivation of the Bogoliubov energy
spectrum for the photonikos is given in the same section. Then in
Sec.IV, we derive the Pauli kinetic equation from the Wigner-Moyal
equation. Section V is devoted to BEC. The Fokker-Planck equation
for photonikos, by which we discuss the possibility of the
creation of Bose-Einstein condensate and evaporation of the
photons from the condensate, is obtained in the same section.
Finally, a brief summary and discussion of our results are given
in the last section.

\section{Basic Equations}
If the intensity of photons is sufficiently large, then the
photon-photon interaction can become more likely than the
photon-plasma particle interaction. Under these conditions, which
we shall assume to be realized, we may consider the medium to
consist of two weakly interacting subsystems: the photon gas and
the plasma, which slowly exchange energy between each other. In
other words, relaxation in a photon-plasma system is then a
two-stage process: firstly, the statistical equilibrium is
established in each subsystem independently, with some average
energies $E_\gamma=K_BT_\gamma$ and $E_p=K_BT_p$, where $K_B$ is
the Boltzmann constant, $T_\gamma$ is the characteristic
"temperature" associated with the average kinetic energy of the
photon, and $T_p$ is the plasma temperature. The photon effective
"temperature",in general, will differ from the plasma temperature.
Slower processes of the equalization of the photon and the plasma
temperatures will take place afterwards.

In the following we will show that under some conditions the
photon-photon interactions dominate the photon-particle
interactions. When the photon-photon interaction takes place, the
phases of the waves are, in general, random functions of time. We
need therefore not be interested in the phases and can average
over them. In such a situation the perturbation state of the
photon gas can be described in terms of the occupation number
$N(\vec{k},\omega,\vec{r},t)$ of photons, and one can study how
these numbers change due to the processes of interaction of the
photons with each other, or with plasma electrons. Note that
$N(\vec{k},\omega,\vec{r},t)$ is the slowly varying function in
space and time.

Recently, a new version of kinetic equation for the occupation
number $N(\vec{k},\omega,\vec{r},t)$ of photons, for modes
propagating with the wavevector $\vec{k}$ and the frequency
$\omega$, at the position $\vec{r}$ and time t, was derived in
Refs.\cite{ltsin98},\cite{ntsin98},\cite{men}. It should be
emphasized that the previous derivations of kinetic equation were
based on envelope equations, were restricted to nonrelativistic
plasmas and neglected the time variation of the photon (or the
wavepacket) mean frequency. In contrast, our derivation avoids the
envelope approximation, is valid for a fully relativistic plasma,
and includes space and time correlations. Since this equation has
appeared previously in the literature
\cite{ltsin98},\cite{ntsin98},\cite{men}, it is presented here
without derivation
\begin{eqnarray}
\frac{\partial}{\partial
t}N(\vec{k},\omega,\vec{r},t)+\frac{c^2}{\omega}(\vec{k}\cdot\nabla)
N(\vec{k},\omega,\vec{r},t)
-\omega_p^2sin\frac{1}{2}\Bigl(\nabla_{\vec{r}}\cdot
\nabla_{\vec{k}}-\frac{\partial}{\partial
t}\frac{\partial}{\partial\omega}\Bigr)\cdot\rho\frac{
N(\vec{k},\omega,\vec{r},t)}{\omega}=0 \ , \label{kin}
\end{eqnarray}
where $\omega_p=\sqrt{\frac{4\pi e^2n_{0e}}{m_{0e}}}$,
$\rho=\frac{n_e}{n_{0e}}\frac{1}{\gamma}$, $m_{0e}$ is the
electron rest mass, $n_e$ and $n_{0e}$ are the non-equilibrium and
equilibrium densities of the electrons, respectively, and $\gamma$
is the relativistic gamma factor of the electrons, which can be
expressed as
\begin{eqnarray}
\gamma=\sqrt{1+Q}=\sqrt{1+\beta\int\frac{d\vec{k}}{(2\pi)^3}\int\frac{d\omega}{2\pi}\frac{
N(\vec{k},\omega,\vec{r},t)}{\omega}} \ , \label{gam}
\end{eqnarray}
where $\beta=\frac{2\hbar\omega_p^2}{m_{0e}n_{0e}c^2}$ and $\hbar$
is the Planck constant divided by $2\pi$.

Equation (\ref{kin}) is the generalization of the Wigner-Moyal
equation for the classical electromagnetic (EM) field. We
specifically note that from this equation follows conservation of
the total number of photons, but not the momentum and energy of
photons, i.e.,
\begin{eqnarray}
N=2\int
d\vec{r}\int\frac{d\vec{k}}{(2\pi)^3}\int\frac{d\omega}{2\pi}
N(\vec{k},\omega,\vec{r},t)=const. \ , \label{con}
\end{eqnarray}
where coefficient 2 denotes two  possible polarization of the
photons. Hence, the chemical potential, $\mu_\gamma$, of the
photon gas is not zero.

If the local frequency $\omega$ and wavevector $\vec{k}$ are
related by the dispersion equation, $\omega=\omega (k)$, then the
occupation number is represented as
$N(\vec{k},\omega,\vec{r},t)=2\pi N(\vec{k},\vec{r},t)\delta
(\omega - \omega (k))$ and Eq.(\ref{kin}) reduces to
\begin{eqnarray}
\frac{\partial}{\partial
t}N(\vec{k},\vec{r},t)+c^2\vec{k}\nabla\frac{
N(\vec{k},\vec{r},t)}{\omega}
-\omega_p^2sin\frac{1}{2}(\nabla_{\vec{r}}\cdot
\nabla_{\vec{k}})\cdot\rho\frac{N(\vec{k},\vec{r},t)}{\omega}=0 \
. \label{kinom}
\end{eqnarray}

In the geometric optics approximation ($sinx\approx x$),
Eq.(\ref{kin}) reduces to the one-particle Liouville-Vlasov
equation with an additional term
\begin{eqnarray}
\frac{\partial}{\partial
t}N(\vec{k},\omega,\vec{r},t)+\frac{c^2}{\omega}(\vec{k}\cdot\nabla)
N(\vec{k},\omega,\vec{r},t)
-\frac{\omega_p^2}{2}\Bigl(\nabla_{\vec{r}}\rho\cdot\nabla_{\vec{k}}-\frac{
\partial\rho}{\partial
t}\cdot\frac{\partial}{\partial\omega}\Bigr)\frac{N(\vec{k},\omega,\vec{r},t)}
{\omega}=0
\ . \label{liou}
\end{eqnarray}

From this kinetic equation we can obtain a set of fluid equations.
To this end, we introduce the definitions of mean values of the
density, velocity and $Q=\gamma^2-1$ of photons:
\begin{eqnarray}
n_\gamma(\vec{r},t)=2\int
\frac{d^3k}{(2\pi)^3}\int\frac{d\omega}{2\pi}N(\vec{k},\omega,\vec{r},t)\
. \label{den}
\end{eqnarray}
The photon mean velocity is defined by
\begin{eqnarray}
\vec{u}_\gamma(\vec{r},t)=\frac{2}{n_\gamma}\int\frac{d^3k}{(2\pi)^3}\int
\frac{d\omega}{2\pi}\frac{\vec{k}c^2}{\omega}N(\vec{k},\omega,\vec{r},t)\
, \label{vel}
\end{eqnarray}
and
\begin{eqnarray}
Q=\frac{\hbar\omega_p}{m_{0e}n_{0e}c^2}2\omega_p\int\frac{d^3k}{(2\pi)^3}\int
\frac{d\omega}{2\pi}\frac{N(\vec{k},\omega,\vec{r},t)}{\omega}\ .
\label{qu}
\end{eqnarray}

Having these definitions, we shall construct the transport
equations by the usual way. Namely, multiplying Eq.(\ref{liou}) by
1, $\frac{\vec{k}c^2}{\omega}$, $\frac{\hbar}{\omega}$ and
integrating over the entire spectral range, in $\vec{k}$ and
$\omega$, we obtain equations of continuity, motion and $Q$ of the
photon gas:
\begin{eqnarray}
\frac{\partial n_\gamma}{\partial t}+div n_\gamma\vec{u}_\gamma=0
\ , \label{conti}
\end{eqnarray}
\begin{eqnarray}
\frac{\partial \vec{u}_\gamma}{\partial
t}+(\vec{u}_\gamma\cdot\vec{\nabla})\vec{u}_\gamma=-\frac{1}{n_\gamma}\nabla
P_\gamma-\frac{U}{2n_\gamma}\Bigl(\nabla\rho+\frac{\vec{u}}{c^2}\cdot
\frac{\partial \rho}{\partial t}\Bigr)\ , \label{motion}
\end{eqnarray}
\begin{eqnarray}
\frac{\partial Q}{\partial t}+div Qu_\gamma=G\frac{\partial
\rho}{\partial t}\ , \label{equ}
\end{eqnarray}
where we have introduced the following notations in dimensional
units
\begin{eqnarray}
U=\omega_p^2c^22\int\frac{d^3k}{(2\pi)^3}\int\frac{d\omega}{2\pi}\frac{
N(\vec{k},\omega,\vec{r},t)}{\omega^2}\ , \label{notu}
\end{eqnarray}
\begin{eqnarray}
P_\gamma =\frac{2}{3}
\int\frac{d^3k}{(2\pi)^3}\int\frac{d\omega}{2\pi}\Bigl(\frac{\vec{k}c^2}{
\omega}-\vec{u}_\gamma\Bigr)^2N(\vec{k},\omega,\vec{r},t)=n_\gamma
T_\gamma\ . \label{notP}
\end{eqnarray}

Obviously, $P_\gamma$ is the pressure of the photon gas in unit
mass. For G we have the following expression
\begin{eqnarray}
G=\frac{\hbar\omega_p}{m_{0e}c^2}\frac{\omega_p^2}{n_{0e}}\cdot 2
\int\frac{d^3k}{(2\pi)^3}\int\frac{d\omega}{2\pi}\frac{
N(\vec{k},\omega,\vec{r},t)}{\omega^3}\approx
\frac{\hbar\omega_p}{m_{0e}c^2}\frac{\omega_p^3}{<\omega^3>}\cdot
\frac{n_\gamma}{n_{0e}}\ . \label{gexp}
\end{eqnarray}

From Eq.(\ref{equ}) follows that the term on the right-hand side
(RHS) is much less than the first term on the left-hand side
(LFS), if $G/\gamma^3\ll 1$. This inequality allows us to neglect
the term on RHS in Eq.(\ref{equ}). With this assumption from
Eqs.(\ref{conti}) and (\ref{equ}) follows the frozen-in condition
\begin{eqnarray}
\frac{Q}{n_\gamma}=const \ . \label{froz}
\end{eqnarray}

It is important to emphasize that in Eqs.(\ref{kin}),
(\ref{kinom}), (\ref{liou}) and (\ref{motion}) there are two
forces of distinct nature, which can change the occupation number
of photons. One force appears due to the redistribution of
electrons in space, $\nabla n_e$, and time, $\frac{\partial
n_e}{\partial t}$. The other force arises by the variation of the
shape of wavepacket. In other words, this force originates from
the alteration of the average kinetic energy of the electron
oscillating in a rapidly varying field of the EM waves, and is
proportional to $\vec{\nabla}\gamma$ and $\frac{\partial}{\partial
t}\frac{1}{\gamma}$.

\section{Stability of the photon flow}

In order to study the problem of stability of a photon flow, we
consider the propagation of small perturbations in a homogeneous
photon flux-plasma medium. First, we will derive a relation
between the variation of photon and plasma densities, from which
we will establish the condition that allows us to neglect the
variation of the plasma density in comparison with the variation
of the photon density.

To show this, we suppose the temperature of electrons to be
nonrelativistic, i.e., $T_e\ll m_{0e}c^2\gamma$.  In this case,
the equations describing a state of the electrons are as follows
\begin{eqnarray}
\frac{\partial\vec{p}_e}{\partial
t}+(\vec{v}_e\cdot\vec{\nabla})\vec{p}_e=-e\vec{E}-m_{0e}c^2\nabla\gamma
\ , \label{elmot}
\end{eqnarray}
\begin{eqnarray}
\frac{\partial n_e}{\partial
t}+div\frac{\vec{p}_e}{m_{0e}\gamma}n_e=0\ .\label{elconti}
\end{eqnarray}

We linearize Eqs.(\ref{elmot}) and (\ref{elconti}) with respect to
perturbations, which are represented as
$\vec{p}_e=\delta\vec{p}_e$, $n_e=n_{0e}+\delta n_e$ and
$\gamma=\gamma_0+\delta\gamma=\gamma_0+\frac{\delta
Q}{2\gamma_0}$, where the suffix $0$ denotes the constant
equilibrium value, and $\delta\vec{p}_e$, $\delta n_e$, and
$\delta Q$ are small variations in the wave. Taking into account
the frozen-in condition (\ref{froz}), $\delta Q$ can be expressed
by the variation of the photon density as $\delta
Q=Q_0\frac{\delta
n_\gamma}{n_{0\gamma}}=\frac{\gamma_0^2-1}{n_{0\gamma}}\delta
n_\gamma$. After linearization, we will seek plane wave solutions
proportional to $exp i(\vec{q}\vec{r} -\Omega t)$.

Here we consider the range of low frequencies for which the
inequalities $\Omega<qc<\omega_{pe}$ are fulfilled. In this case
the photon flow can no longer excite the Langmiur plasma waves,
and the contribution of perturbation of the electron density is
rather small in comparison to perturbation of the photon density
as
\begin{eqnarray}
\frac{\delta
n_e}{n_{0e}}=-\frac{\gamma_0^2-1}{2\gamma_0^2}\frac{c^2q^2}
{\omega_{pe}^2}\frac{\delta n_\gamma}{n_{0\gamma}} \ .
\label{deltarel}
\end{eqnarray}
Therefore, in our consideration the density of the plasma
particles remain constant. In this case Eqs. (\ref{kin}) and
(\ref{kinom}) contain the variation of the relativistic $\gamma$
factor alone.

In the following we shall demonstrate a new phenomena in the
photon gas, which originates from the variation of shape of the
wavepacket. Obviously, any waves, which can arise in the photon
gas-plasma medium in this case, are the proper waves with the weak
decrement of the photon gas. It means that these waves are
associated with the photon gas, but not with the plasma.

To this end, we linearize Eq.(\ref{kinom}) with respect to the
perturbation, which is represented as
\begin{eqnarray}
N(\vec{k},\omega(\vec{k}),\vec{r},t)=N_0(\vec{k},\omega(\vec{k}))+\delta
N(\vec{k},\omega(\vec{k}),\vec{r},t)\ .\label{perturb}
\end{eqnarray}
The result is
\begin{eqnarray}
\Bigl(\frac{\vec{q}\vec{k}c^2}{\omega}-\Omega\Bigr)\delta
N=-\omega_p^2\frac{\delta\gamma}{\gamma_0^2}\sum_{l=0}^\infty\frac{(\vec{q}
\cdot\nabla_{\vec{k}})^{2l+1}}{(2l+1)!2^{2l+1}}\cdot\frac{N_0(\vec{k},\omega(k))}{
\omega(k)} \ , \label{linkin}
\end{eqnarray}
or after summation we obtain
\begin{eqnarray}
\Bigl(\frac{\vec{q}\vec{k}c^2}{\omega}-\Omega\Bigr)\delta
N=-\omega_p^2\frac{\delta\gamma}{\gamma_0^2}\Bigl\{\frac{N_0^+(\vec{k}+\vec{q}/2)}{
\omega(\vec{k}+\vec{q}/2)}-\frac{N_0^-(\vec{k}-\vec{q}/2)}{
\omega(\vec{k}-\vec{q}/2)}\Bigr\} \ , \label{sumkin}
\end{eqnarray}
where $\delta\gamma=\frac{1}{2\gamma_0}\delta
Q=\frac{\beta}{2\gamma_0}\int\frac{d^3k}{(2\pi)^3}\cdot\frac{\delta
N(k,\omega,q,\Omega)}{\omega(k)}.$

After integration over all wavevectors $\vec{k}$, from
Eq.(\ref{sumkin}) we get the dispersion relation due to the
relativistic selfmodulation
\begin{eqnarray}
1+\frac{\omega_p^2}{2\gamma_0^3}\beta\int d^3k\Bigl\{
\frac{N_0^+(\vec{k}+\vec{q}/2)}{
\omega(\vec{k}+\vec{q}/2)}-\frac{N_0^-(\vec{k}-\vec{q}/2)}{
\omega(\vec{k}-\vec{q}/2)}\Bigr\}/\omega(k)
\Bigl(\frac{\vec{q}\vec{k}c^2}{\omega(k)}-\Omega\Bigr)=0 \ ,
\label{self}
\end{eqnarray}
which has, in general, complex roots.

This equation has been derived and predicted by us first for the
monochromatic wave in Ref.\cite{ltsin91}, and then for the broad
spectrum in Ref.\cite{ltsin98}.

In Eq.(\ref{self}) $N_0(k)$ is the occupation number in the
equilibrium state and is represented as
\begin{eqnarray}
N_0(k)=n_{0\gamma}(2\pi\sigma_k^2)^{-3/2}exp\Bigl(-\frac{(\vec{k}-\vec{k}_0)^2}{
2\sigma_k^2}\Bigr) \ .\label{gauss}
\end{eqnarray}
This is a spectral Gaussian distribution, with the average
wavevector $\vec{k}_0$ and the spectral width $\sigma_k$.

We can rewrite Eq.(\ref{self}) in another form, taking into
account a pole in the integral
\begin{eqnarray}
\Omega-\vec{q}\vec{u}=0 \ , \label{pole}
\end{eqnarray}
where $\vec{u}=\frac{\vec{k}c^2}{\omega(k)}<c$.

Using the well known relation
\begin{eqnarray}
\lim_{\varepsilon\rightarrow 0}\frac{1}{x+\imath\varepsilon}=\wp
\frac{1}{x}-\imath\pi\delta(x) \ , \label{wellk}
\end{eqnarray}
where $\wp$ denotes the prescription that at the singularity $x=0$
the principal value is to be taken, Eq.(\ref{self}) is rewritten
in the form
\begin{eqnarray}
1+\frac{\omega_p^2\beta}{2\gamma_0^3}\wp\int
\frac{d^3k}{\omega(k)}\Bigl\{ \frac{N_0^+(\vec{k}+\vec{q}/2)}{
\omega(\vec{k}+\vec{q}/2)}-\frac{N_0^-(\vec{k}-\vec{q}/2)}{
\omega(\vec{k}-\vec{q}/2)}\Bigr\}/ (\vec{q}\vec{u}-\Omega)\nonumber \\
+\imath \frac{\pi\omega_p^2\beta}{2\gamma_0^3}\int
\frac{d^3k}{\omega(k)}\Bigl\{ \frac{N_0^+(\vec{k}+\vec{q}/2)}{
\omega(\vec{k}+\vec{q}/2)}-\frac{N_0^-(\vec{k}-\vec{q}/2)}{
\omega(\vec{k}-\vec{q}/2)}\Bigr\}\delta(\Omega-\vec{q}\vec{u})=0 \
. \label{selpol}
\end{eqnarray}
In the first and the third integrals we now replace the wavevector
$\vec{k}+\vec{q}/2$ by $\vec{k}$, and in the second and the forth
integrals $\vec{k}-\vec{q}/2\rightarrow\vec{k}$. Assuming that
$\vec{k}=\vec{k}_0+\vec{\chi}$, $\mid q\mid ,$
$\mid\vec{\chi}\mid\ll\mid\vec{k}_0\mid$, after integration we
obtain the dispersion relation
\begin{eqnarray}
1+\frac{q^2V_E^2}{(\Omega-\vec{q}\vec{u}_g)^2-q^2V_s^2-\alpha^2q^4}+\frac{
\imath\pi\omega_p^2\beta}{2\gamma_0^3c^2\omega(k_0)}\Bigl(\frac{\partial
N_0(\chi_z)}{\partial
\chi_z}\Bigr)_{\chi_z=\frac{\omega(k_0)}{qc^2}(\Omega-\vec{q}\vec{u}_g)}=0\
, \label{disp}
\end{eqnarray}
where $\vec{u}_g=\frac{\vec{k}_0c^2}{\omega(k_0)}$,
$\vec{q}\vec{u}_g=qu_g\cos\Theta$, $N_0(\chi_z)=\int d\chi_x\int
d\chi_yN_0(\chi), $
$V_E=c\frac{\omega_p}{\omega(k_0)}\frac{1}{\sqrt{2\gamma_0}}\sqrt{\frac{\gamma_0^2-1}{
\gamma_0^2}}$, $\alpha=\frac{c^2}{2\omega(k_0)}$, and the velocity
$V_s=c\sqrt{3}\sqrt{\frac{\gamma_0^2-1}{
\gamma_0^2}}\frac{\sigma_k c}{\omega(k_0)}$ can be treated as the
sound velocity of longitudinal photons, similar to phonons in the
quantum liquid at almost zero temperature \cite{lif}.

We now first neglect the small imaginary term in Eq.(\ref{disp}),
and examine in great detail the following equation
\begin{eqnarray}
(\Omega-\vec{q}\vec{u}_g)^2=q^2\Bigl\{V_s^2+\alpha^2q^2-V_E^2\Bigr\}\
. \label{ditdis}
\end{eqnarray}
This equation has a complex solutions, when the inequality
\begin{eqnarray}
V_E^2>V_s^2+\alpha^2q^2 \ ,\label{vevs}
\end{eqnarray}
for $q<\frac{2\omega(k)}{c^2}\sqrt{V_E^2-V_s^2}$, holds. The
unstable solution can be written as
$\Omega=\Omega\prime+\imath\Omega\prime\prime$, where
\begin{eqnarray}
\Omega\prime=qu_g\cos\Theta  \label{reom}
\end{eqnarray}
and the growth rate for the unstable modes with
$q<\frac{2\omega(k)}{c^2}\sqrt{V_E^2-V_s^2}$ is
\begin{eqnarray}
\Omega\prime\prime=q\sqrt{V_E^2-V_s^2-\alpha^2q^2 } \ .
\label{imom}
\end{eqnarray}

This instability for the monochromatic relativistic EM waves was
predicted in Ref.\cite{ltsin91}, and for the broad spectrum at
$V_s=\alpha=0$ was disclosed in Ref.\cite{ntsin98}.

If $\Omega\prime>\mid\Omega\prime\prime\mid$, then the solution
(\ref{reom}) clearly describes the emission of the longitudinal
photons (photonikos) by the bunch of the transverse photons
(photons) inside a resonance cone
$(\cos\Theta=\Omega\prime/qu_g)$, similar to the well-known
Cherenkov emission of EM waves by a charged particles moving in
uniform medium with  a velocity larger than the phase velocities
of emitted waves. Also, here is an analogy with the Landau
criterion about the creation of elementary excitations.

From Eq.(\ref{imom}) follows that there exists two range of values
of the wavevector q. Namely, for all q less than
$\frac{2\omega}{c^2}\sqrt{V_E^2-V_s^2}$, the Cherenkov type
mechanism leads to the excitation of  photonikos (large scale
waves). For the other, q more than
$\frac{2\omega}{c^2}\sqrt{V_E^2-V_s^2}$, the frequency
$\Omega=\Omega\prime+\imath\Omega\prime\prime$ is real.

We now consider the case, when $V_s^2>V_E^2$ or
$\omega_p^2<\gamma_0\sigma_k^2c^2$. If we write Eq.(\ref{ditdis})
in the moving frame ($\Omega-\vec{q}\vec{u}\rightarrow\Omega$),
and multiply the both sides by the Planck constant $\hbar$, then
we obtain the well known Bogoliubov energy spectrum
\begin{eqnarray}
\varepsilon(p)=\sqrt{V^2p^2+\Bigl(\frac{p^2}{2m_{eff}}\Bigr)^2} \
,\label{bogol}
\end{eqnarray}
where $\varepsilon=\hbar\Omega$, $p=\hbar q$, and
$m_{eff}=\hbar\omega/c^2$.

Equation (\ref{bogol}) has the same form as the energy of the
elementary excitations in the quantum liquid at zero temperature,
which was derived first by Bogoliubov.

For small momentum $(p<m_{eff}V=m_{eff}\sqrt{V_s^2-V_E^2})$ from
Eq.(\ref{bogol}) follows $\varepsilon=pV\approx V_sp.$ This
expression show us that the coefficient V is the velocity of
"sound" in the photon gas, similar to phonons in the quantum
medium. For the large momentum ($p\gg m_{eff}V$), the photoniko
energy (\ref{bogol}) tends to $p^2/2m_{eff}$, i.e., the kinetic
energy of an individual photoniko.

Thus, we may conclude that the elementary excitations in our case
correspond to photonikos.

The decrement, $Im\Omega=\Omega\prime\prime$, of photoniko is
derived from Eq.(\ref{disp}), and the result is
\begin{eqnarray}
Im\Omega=-\frac{\omega_p}{\gamma_0^4}F\Bigl(\frac{\gamma_0^2-1}{\gamma_0^2}\Bigr)^2
exp\Bigl\{-\frac{(\Omega-\vec{q}\vec{u})^2}{2q^2v_0^2}\Bigr\} \
,\label{decr}
\end{eqnarray}
where $v_0^2=c^2\frac{\sigma_k^2c^2}{\omega^2(k_0)}$,
$F=\sqrt{\frac{\pi}{32}}\Bigl(\frac{\omega_p}{\omega(k_0)}\Bigr)
\Bigl(\frac{\omega_p}{\sigma_kc}\Bigr)^2\frac{q}{\sigma_k}.$

Clearly the expression (\ref{decr}) is valid for the weakly damped
oscillations, that is $Re\Omega\gg Im\Omega$. We note that the
decrement has a maximum near the frequency of the Cherenkov
resonance $\Omega\simeq qu\cos\Theta$.

\section{Pauli equation for the photon gas}

 In the previous section we have derived the condition
(\ref{deltarel}) allowing the situation, in which the density of
plasma almost does not change, to exist. In this case from
Eq.(\ref{kin}) follows that photons with different frequencies and
wavevectors scatter on the photon bunch (wavepacket), and the
equilibrium state may be established. This is a new phenomena of
the "Compton" scattering type, i.e., the statistical equilibrium
between the photons and the photon bunch will establish itself as
a result of the scattering processes.

In order to better exhibit this, we shall derive the Pauli
equation in the wavevector representation. Note that the
Wigner-Moyal-Tsintsadze equation (\ref{kin}) is convenient for a
study of non-linear processes in the photon gas in the case of
so-called weak turbulence states. Such states of the photon
gas-plasma means that the energy of the quasi-particle (photoniko)
is small compared to the photon energies.

Usually the weak-turbulence states arise in a medium, when a small
perturbation grows with the growth rate smaller than the
frequency, $Im\Omega\ll Re\Omega$. In this case we can use a
series expansions in powers of amplitude of  the perturbations.

In the case of weak-turbulence, we can define the distribution
function $N(\vec{k},\omega,\vec{r},t)$ in any approximation for
the amplitude of perturbation by iterating the Wigner-Moyal
equation. From Eq.(\ref{kin}), we now derive an equation in
wavevector space in the limit of the spatial homogeneity, which
may be identified with the Pauli equation. To this end, we take
the Fourier transform of Eq.(\ref{kin}) to obtain
\begin{eqnarray}
\frac{\partial}{\partial
t}N(\vec{k},\omega,\vec{q},\Omega,t)-\imath\Bigl(\Omega-\frac{\vec{q}\vec{k}c^2}{\omega}\Bigr)
N(\vec{k},\omega,\vec{q},\Omega,t)\nonumber \\
=\omega_p^2\int\frac{
d^3q\prime}{(2\pi)^3}\int\frac{d\Omega\prime}{2\pi}\delta\rho(\vec{q}\prime,
\Omega\prime)\sin\frac{\imath}{2}\Bigl(\vec{q}\prime\cdot\nabla_{\vec{k}}+
\Omega\prime\frac{\partial}{\partial\omega}\Bigr)\frac{
N(\vec{k},\omega,\vec{q}-\vec{q}\prime,\Omega-\Omega\prime,t)}{\omega}
\ , \label{four}
\end{eqnarray}
where $N(\vec{k},\omega,\vec{q},\Omega,t)=\int d\vec{r}\int dt
N(\vec{k},\omega,\vec{r},t)exp\Bigl(-\imath(\vec{q}\cdot\vec{r}-\Omega
t)\Bigr),$ and $\delta\rho(\vec{q},\Omega,t)=\int d\vec{r}\int dt
\delta\rho(\vec{r},t)exp\Bigl(-\imath(\vec{q}\cdot\vec{r}-\Omega
t)\Bigr)$ are the slowly varying functions in time.

Let us consider two cases. First, in the case of $\vec{q}=0$ and
$\Omega=0$, after changing variables $-\vec{q}\prime\rightarrow
\vec{q}\prime$ and $-\Omega\prime\rightarrow  \Omega\prime$, we
rewrite Eq.(\ref{four}) in the form
\begin{eqnarray}
\frac{\partial N(\vec{k},\omega,t)}{\partial
t}=-\omega_p^2\int\frac{
d^3q}{(2\pi)^3}\int\frac{d\Omega}{2\pi}\delta\rho(-\vec{q},
-\Omega,t)\hat{L}\frac{
N(\vec{k},\omega,\vec{q},\Omega,t)}{\omega(k)} \ . \label{first}
\end{eqnarray}
Here we have used the identity
\begin{eqnarray}
\sin\frac{\imath}{2}\Bigl(\vec{q}\cdot\nabla_{\vec{k}}+
\Omega\frac{\partial}{\partial\omega}\Bigr)=
\imath\sinh\frac{1}{2}\Bigl(\vec{q}\cdot\nabla_{\vec{k}}+
\Omega\frac{\partial}{\partial\omega}\Bigr)=\imath\hat{L} \ .
\label{iden}
\end{eqnarray}

For the second case, when $\vec{q}\neq 0$ and $\Omega\neq 0$ (in
this case $\partial\ln N/\partial t\ll\Omega$), Eq.(\ref{four})
can be written as
\begin{eqnarray}
\Bigl(\Omega-\frac{\vec{q}\vec{k}c^2}{\omega}\Bigr)
N(\vec{k},\omega,\vec{q},\Omega)=-\omega_p^2\int\frac{
d^3q\prime}{(2\pi)^3}\int\frac{d\Omega\prime}{2\pi}\delta\rho(-\vec{q}\prime,
-\Omega\prime)\hat{L}\frac{
N(\vec{k},\omega,\vec{q}-\vec{q}\prime,\Omega-\Omega\prime)}{\omega}
\ . \label{second}
\end{eqnarray}
By iterating this equation in the case of weak-turbulence we can
write an equation for the occupation number N in any approximation
in the amplitude of perturbations. Expanding the integrand about
$\vec{q}=\vec{q}\prime$, $\Omega=\Omega\prime$ and keeping the
first term with
$N(\vec{k},\omega,\vec{q}-\vec{q}\prime,\Omega-\Omega\prime)=(2\pi)^4\delta
(\vec{q}\prime-\vec{q})\delta
(\Omega\prime-\Omega)N(\vec{k},\omega),$ we obtain
\begin{eqnarray}
N(\vec{k},\omega,\vec{q},\Omega)=-\omega_p^2\frac{\delta\rho(\vec{q},
\Omega)\hat{L}\frac{N(\vec{k},\omega)}{\omega}}{\Omega-\vec{q}\cdot\vec{u}}
\ .\label{secturb}
\end{eqnarray}

Substituting Eq.(\ref{secturb}) into Eq.(\ref{first}) we get
\begin{eqnarray}
\frac{\partial N(\vec{k},\omega)}{\partial
t}=\imath\omega_p^4\int\frac{
d^3q}{(2\pi)^3}\int\frac{d\Omega}{2\pi}\hat{L}\frac{\mid\delta\rho(\vec{q},
\Omega)\mid^2}{\omega(\Omega-\vec{q}\cdot\vec{u})}\hat{L}\frac{
N(\vec{k},\omega)}{\omega} \ . \label{firnew}
\end{eqnarray}

Since we are considering the weak-turbulence, for the perturbation
the dispersion relation $\Omega=\Omega(q)$ is valid, i.e.,
\begin{eqnarray}
\delta\rho(\vec{q},\Omega)=2\pi\delta(\Omega-\Omega(q))\delta\rho(\vec{q})
\ . \label{weakdis}
\end{eqnarray}
Noting the relation
\begin{eqnarray}
\sinh(y\frac{\partial}{\partial
x})f(x)=\frac{1}{2}\Bigl\{f(x+y)-f(x-y)\Bigr\} \label{sinh}
\end{eqnarray}
and $\delta\rho(-\vec{q},-\Omega)=\delta\rho^\ast(\vec{q},\Omega)$
(where the asterisk $^\ast$ depicts the Hermitian conjugate), we
obtain from Eq.(\ref{firnew})
\begin{eqnarray}
\frac{\partial N(\vec{k},\omega)}{\partial
t}=\imath\frac{\omega_p^4}{4}\int\frac{
d^3q}{(2\pi)^3}\frac{\mid\delta\rho(q)\mid^2}{
\omega(\vec{k}+\vec{q}/2)}\Bigl\{\frac{
N(\vec{k}+\vec{q},\omega+\Omega)}{\omega(\vec{k}+\vec{q})} -
\frac{ N(\vec{k},\omega)}{\omega(\vec{k})}\Bigr\}\nonumber \\
\times\Bigl[\frac{1}{ \Omega-\vec{q}\cdot\vec{u}\prime}+\frac{1}{
\Omega+\vec{q}\cdot\vec{u}\prime}\Bigr] \ , \label{herm}
\end{eqnarray}
where
$\vec{u}\prime=\frac{(\vec{k}+\vec{q}/2)c^2}{\omega(\vec{k}+\vec{q}/2)}$,
and use was made of dispersion relation $\omega=\omega(k)$.

Use of Eq.(\ref{wellk}) in Eq.(\ref{herm}) yields the following
relation
\begin{eqnarray}
\frac{\partial N(\vec{k})}{\partial t}=\sum_{\pm}\int\frac{
d^3q}{(2\pi)^3}\Bigl[W_\pm(\vec{k}+\vec{q},\vec{k})N(\vec{k}+\vec{q})-
W_\pm(\vec{k},\vec{k}+\vec{q})N(\vec{k})\Bigr] \ .\label{fineq}
\end{eqnarray}

Here we have introduced the scattering rate
\begin{eqnarray}
W_\pm=\frac{\pi}{2}\frac{\omega_p^4\mid\delta\rho(q)\mid^2}{
\omega(\vec{k}+\vec{q}/2)\omega(k)}\delta(\Omega\pm\vec{q}\cdot\vec{u}\prime)\
.\label{scatt}
\end{eqnarray}

Equation (\ref{fineq}) describes the three wave interaction, as
illustrated in Fig.1. First (a) diagram exhibits the absorption of
the photoniko by the photon, whereas the second (b) diagram shows
the emission of the photoniko by the photon. In other words, the
photon passing through the photon bunch (wavepacket) absorbs and
emits photonikos, with frequencies
$\Omega=\mp(\omega-\omega\prime)$ and wavevectors
$\vec{q}=\mp(\vec{k}-\vec{k}\prime)$.

Integral in Eq.(\ref{fineq}) is the elastic collision integral and
describes the photon scattering process on the variation of shape
of the photon bunch.

It should be emphasized that the kinetic equation (\ref{fineq}) is
a new version of the Pauli equation in the wavevector
representation \cite{pau}. Note that the equation type of
(\ref{fineq}) has been obtained by Pauli for a quantum system,
whereas Eq.(\ref{fineq}), derived for the dense photon gas, is
pure classical. This equation indicates that the equilibrium of
the photon gas is triggered by the perturbation $\delta\rho=\delta
(n_/n_o\gamma)$, in particular by the perturbation in the
wavepacket.

\section{Bose-Einstein condensation in the photon gas}

As we have mentioned in the Introduction, Zel'dovich and Levich
have shown that the usual Compton scattering leads to the BEC in
the nonrelativistic photon gas-plasma medium.

In this section, we will show that exists an another new mechanism
of BEC in a non-ideal dense photon gas. As was mentioned above, if
the intensity of radiation  is sufficiently large, then the
photon-photon interaction can become more likely than the
photon-electron interaction.

It is well known from plasma physics that the transverse photon
can decay into the transverse photon and a plasmon (electron
Langmuir and ion sound plasmons). For this process to take place
it is necessary to have a fluctuation of the plasma density.

In our consideration, as was shown in Sec.III, the density of
plasma remains constant. So that only one mechanism (relativistic
effect), which can cause BEC, is the decay of the transverse
photon into the transverse and the longitudinal (photoniko)
photons. That is, in the weak-turbulence limit this mechanism is
realized by three wave interaction, as we have discussed in
Sec.IV. Namely, the photon ($\omega,\vec{k})$ generates photoniko
and scatters on latter, with the frequency $\omega\prime$ and the
wavevector $\vec{k}\prime$. In this case, the wave number and
frequency matching conditions are satisfied, i.e.,
$\vec{k}=\vec{k}\prime+\vec{q}$, and $\omega=\omega\prime+\Omega$.
These processes continue as a cascade
$(\omega\prime=\omega\prime\prime+\Omega\prime,\
\vec{k}\prime=\vec{k}\prime\prime+\vec{q}\prime$, etc.) till the
wavevector of the photon becomes zero and the frequency
$\omega=\omega_p/\gamma^{1/2}$. This cascade leads to the BEC of
the photons and the creation of the photoniko gas. When the
density of latter is sufficiently small, the photonikos may be
regarded as non-interacting with each other, however they can
exchange energies and momentum with a bunch of the photons by
scattering mechanism, and finally such process can lead to the
equilibrium state of the photoniko gas, with the Bose distribution
\begin{eqnarray}
n_L=\frac{1}{exp\Bigl\{\frac{\hbar\Omega-\vec{p}\vec{u}}{T_\gamma}\Bigr\}-1}
\ ,\label{photoniko}
\end{eqnarray}
where $\Omega$ is the Bogoliubov spectrum (\ref{bogol}).

We note here that since the total number of photonikos does not
conserve, the chemical potential of this gas is zero.

The result of BEC is that the ground state is filled by photons
with the total rest energy
$E_0=N_0\frac{\hbar\omega_p}{\gamma^{1/2}}$, where $N_0$ is the
total number of photons  in the ground state.

We now express explicitly the energy of the ground state, $E_0$,
through the total number, $N_0$, and the volume of the photon gas
in two limits. First, we consider ultrarelativistic case, when
$\gamma=\sqrt{1+Q}\approx
Q^{1/2}=(\hbar\bar{\omega_p}/m_0c^2)^{2/3}(N_0/N_{0e})^{2/3}(1/V^{1/3})$,
where $\bar{\omega_p}=(4\pi e^2N_{0e}/m_{0e})^{1/2}$, $N_{0e}$ is
the total number of electrons.

The chemical potential of the photon gas is
\begin{eqnarray}
\mu=\frac{\partial E_0}{\partial
N_0}=\frac{2}{3}\Bigl(\frac{N_{0e}}{N_0}\Bigr)^{1/3}(\hbar\omega_p)^{2/3}
(m_0c^2)^{1/3} \ .\label{chempo}
\end{eqnarray}
Note that in quantum Bose liquid at $T=0$, the chemical potential
increases with increase of $N_0$, whereas in the present case
$\mu$ decreases as shows Eq.(\ref{chempo}).

We now find the pressure in the condensate, $P_c$, as
$P_c=-\partial E_0/\partial V$, and write the equation of state
for given $N_0$, the result is
\begin{eqnarray}
P_cV^{4/3}=const\ .\label{staterel}
\end{eqnarray}

Next in the nonrelativistic limit, we have $\gamma\simeq
1+(1/2)(\hbar\bar{\omega_p}/m_0c^2)(N_0/N_{0e})(1/V^{1/2})$. Then
for the equation of state and the chemical potential we get the
following expressions
\begin{eqnarray}
P_cV^{3/2}=const \label{statenon}
\end{eqnarray}
and
\begin{eqnarray}
\mu=\hbar\omega_p\Bigl(1-\frac{1}{2}\frac{N_{0}}{N_{0e}}\frac{
\hbar\omega_p}{m_0c^2}\Bigr) \ .\label{chemponon}
\end{eqnarray}
The second term in Eq.(\ref{chemponon}) is always less than one.

We note that when the conditions for the formation of BEC arise,
then the photon gas in plasmas can be far from the equilibrium.
The kinetics of this phenomena is a very important issue not only
in the study of the contemporary problems of short pulse
laser-matter interaction, but also for understanding of the
processes, which take place on cosmic objects in the presence of
strong radiation of EM field.

The kinetic equation (\ref{fineq}) is the simplest nonequilibrium
model in which we can expect BEC. To show this, we introduce the
number density of the photons in the condensate as
$n_0=\lim_{V\rightarrow\infty}\frac{N_0}{V}$. Then we can write
$N(\vec{k},t)=4\pi^3\delta(\vec{k})n_0(t)$, which means that the
photons with a zero wavevector are in the condensate.

We next discuss an another situation in which the photon with the
wavevector $\vec{k}$ is collided with the photoniko with
$\vec{q}$. For such processes exists the probability of the
condition $\vec{k}+\vec{q}=0$ to be satisfied for the photon and
photoniko. Thus the photons absorbing the photoniko will pass to
the ground state. Therefore, the occupation number
$N(\vec{k}+\vec{q})$, which is under integral in Eq.(\ref{fineq}),
can be written as
\begin{eqnarray}
N(\vec{k}+\vec{q})=4\pi^3\delta
(\vec{k}+\vec{q})n_1(t)+N(\vec{k}+\vec{q})_{\vec{k}\neq -\vec{q}}\
,\label{ocnum}
\end{eqnarray}
where $n_1(t)$ is the density of pairs, for which the equality
$\vec{k}+\vec{q}=0$ holds.

Substituting Eq.(\ref{ocnum}) into Eq.(\ref{fineq}), for the
Bose-Einstein condensate density $n_0(t)$ we obtain
\begin{eqnarray}
\frac{\partial n_0(t)}{\partial
t}=\frac{n_1(t)-n_0(t)}{\tau_0}+2\int_{\vec{k}\prime\neq
0}\frac{d^3k\prime}{(2\pi)^3}\int\frac{d^3q}{(2\pi)^3}N(k\prime,t)
W(\vec{k}\prime,\vec{k}\prime-\vec{q})\ ,\label{dencon}
\end{eqnarray}
where
$\tau_0^{-1}=\int\frac{d^3q}{(2\pi)^3}W(q)=\frac{\pi}{2}\omega_p^4\int
\frac{d^3q}{(2\pi)^3}\frac{\mid\delta\rho
(q)\mid^2}{\omega(q)\omega(q/2)}\delta
\Bigl(\Omega-\frac{q^2c^2}{2\omega_p}\gamma^{1/2}\Bigr),$ and in
the last term  we have replaced
$\vec{k}+\vec{q}\rightarrow\vec{k}\prime$.

We specifically note here that the first and the third terms in
Eq.(\ref{dencon}) are the sources of production of the
Bose-Einstein condensate, whereas the second term describes the
evaporation of photons from the ground state.

As we can see from Eq.(\ref{dencon}), the stationary solution may
exist, and it is
\begin{eqnarray}
n_0=n_1+\frac{\tau_0}{<\tau>}n_2 \ ,\label{statn0}
\end{eqnarray}
where
$<\tau>^{-1}=\frac{2}{n_2}\int\frac{d^3k\prime}{(2\pi)^3}\frac{
N(k\prime)}{\tau(k\prime)},$ with $
\tau(k\prime)^{-1}=\int\frac{d^3q}{(2\pi)^3}W(\vec{k}\prime,\vec{k}\prime-\vec{q})$.

Equation (\ref{statn0}) has an obvious physical meaning, which is
the following: the departure of photons from the Bose-Einstein
condensate should, under steady-state conditions, be completely
balanced by their arrival from other modes.

The problem of BEC and evaporation of the Bose-Einstein condensate
we can investigate by Fokker-Planck equation, which we shall
derive from the Pauli equation. Here we suppose that the
wavevector $\vec{q}$ and the frequency $\Omega$ of photonikos are
small in comparison with the $\vec{k}$ and $\omega$. This allows
us to use the following expansion in the integrand
\begin{eqnarray}
W(\vec{k}+\vec{q},\vec{k})N(\vec{k}+\vec{q})\approx
W(k)N(k)+\vec{q}\frac{\partial}{\partial\vec{k}}(WN)\mid_{q=0}+\frac{q_\imath
q_\jmath}{2}\frac{\partial^2WN}{\partial k_\imath\partial
k_\jmath}\mid_{q=0}+...\ .\label{expan}
\end{eqnarray}

Substituting this expression into Eq.(\ref{fineq}), we obtain the
Fokker-Planck equation for photons, which describes the slow
change of the occupation number in the wavevector space
\begin{eqnarray}
\frac{\partial N(k)}{\partial t}=\frac{\partial}{\partial
k_\imath}\Bigl\{A_\imath N(k)+\frac{1}{2}\frac{\partial}{\partial
k_\jmath}(D_{\imath\jmath}N(k))\Bigr\}\ ,\label{fp}
\end{eqnarray}
where
\begin{eqnarray}
A_\imath=\int\frac{d^3q}{(2\pi)^3}q_\imath W(q,k)\ ,\label{Ai}
\end{eqnarray}
\begin{eqnarray}
D_{\imath\jmath}=\int\frac{d^3q}{(2\pi)^3}q_\imath q_\jmath
W(q,k)\ .\label{Dij}
\end{eqnarray}
The quantity $A_\imath$ is the dynamic friction coefficient,
whereas $D_{\imath\jmath}$ is the diffusion tensor in the
wavevector space.

Introducing the following definition
$B_\imath=A_\imath+\frac{\partial D_{\imath\jmath}}{2\partial
k_\jmath}$, and noting that the expression on RHS in Eq.(\ref{fp})
is divergent in wavevector space, Eq.(\ref{fp}) can be rewritten
in the form
\begin{eqnarray}
\frac{\partial N}{\partial t}+\frac{\partial}{\partial
k_\imath}\jmath_\imath=0\ ,\label{newfp}
\end{eqnarray}
where $\jmath_\imath=-B_\imath
N-\frac{D_{\imath\jmath}}{2}\frac{\partial N}{\partial k_\jmath}$
is the photon flux density in wavevector space. The fact that the
flux should be zero allowed us to express $B_\imath$ and
$D_{\imath\jmath}$ in terms of one another.

The equilibrium distribution function we choose to be Gaussian
\begin{eqnarray}
N(k)=const\cdot e^{-\frac{k^2}{2\sigma_k^2}}\ .\label{gaus}
\end{eqnarray}

Substituting expression (\ref{gaus}) into the equation
$\vec{\jmath}=0$, we obtain
\begin{eqnarray}
B_\imath\sigma_k^2=\frac{1}{2}k_\jmath D_{\imath\jmath}\
.\label{bidij}
\end{eqnarray}

Finally the transport equation of the photon gas takes the form
\begin{eqnarray}
\frac{\partial N}{\partial t}=\frac{\partial}{\partial
k_\imath}\Bigl\{\frac{D_{\imath\jmath}}{2}\Bigl(\frac{k_\jmath N}{
\sigma_k^2}+\frac{\partial N}{\partial k_\jmath}\Bigr)\Bigr\}\
.\label{trans}
\end{eqnarray}

In order to solve Eq.(\ref{trans}), we consider a simple model
representing the diffusion tensor as
$D_{\imath\jmath}=D_0\delta_{\imath\jmath}$, with $D_0=const$.
Such situation is realized when $u\cdot\cos\Theta\ll qc^2/\omega$.
With this assumption Eq.(\ref{trans}) reduces to
\begin{eqnarray}
\frac{\partial N}{\partial t}=a\frac{\partial}{\partial
\vec{k}}(\vec{k}N)+\frac{D_0}{2}\nabla_{\vec{k}}^2N\ ,
\label{simtr}
\end{eqnarray}
where $a=\frac{D_0}{2\sigma_k^2}$.

To discover the physical meaning of terms on RHS of
Eq.(\ref{simtr}), we consider them separately. First we neglect
the diffusion term in Eq.(\ref{simtr}), and assuming
$\vec{k}(0,0,k)$, we get for the new function $f=kN$
\begin{eqnarray}
\frac{\partial f}{\partial t}-ak\frac{\partial f}{\partial k}=0\ .
\label{firstsim}
\end{eqnarray}

The general solution of this equation is an arbitrary function of
$k_0=ke^{at}$ (where $k_0$ is the initial value of the
wavevector), i.e., $f(k_0)=f(ke^{at})$. This function is constant
on the curves $k_0=ke^{at}$, while at $t\rightarrow\infty$ the
wavevector goes to zero $(k\rightarrow 0)$. The meaning of this is
that the occupation number would tend to peak toward the origin as
\begin{eqnarray}
N(k,t)=\frac{f}{k}=const\cdot e^{at}\ . \label{orig}
\end{eqnarray}

Thus we conclude that the friction effect leads to BEC of the
photon gas.

It should be emphasized that the exponential decay of the
wavevector $k=k_0e^{-at}$ is due to the linearity of the
Fokker-Planck equation. Note that we have neglected all nonlinear
terms in deriving the Pauli equation. The presence of nonlinearity
could saturate the exponential growth of $N(k,t)$. In the above
approximation (\ref{orig}) the condensate formation time, $t_c$,
is defined by the relation $at_c\sim 1$.

Second, we suppose that $W(\vec{q},\vec{k})$ is the even function
with respect to $\vec{q}$. In this case $A=0$, and we have
\begin{eqnarray}
\frac{\partial N}{\partial t}=\frac{D_0}{2}\nabla_{k}^2N\ .
\label{even}
\end{eqnarray}

Assuming that initially all photons are in the ground state with
$k=0$, and the occupation number of the photons is
$N=4\pi^3n_0\delta(\vec{k})$, then the solution of Eq.(\ref{even})
reads
\begin{eqnarray}
N(k,t)=n_0 e^{-\frac{k^2}{2D_0t}}/(2\pi D_0t)^{1/2}\ .
\label{solev}
\end{eqnarray}

In the course of time, the number of photons with $k=0$ decreases
as $t^{-1/2}$. The number of photons in the surrounding wavevector
space rises correspondingly, and initially peaked at the origin is
to be flatten out. Let us determine the mean square wavevector
from origin at time t. From expression (\ref{solev}) we have
\begin{eqnarray}
<k^2>=D_0t\ . \label{sqr}
\end{eqnarray}
Thus, $\sqrt{<k^2>}$ increases as the square root of time.

Eqs. (\ref{solev}) and (\ref{sqr}) manifest that the evaporation
of the photons from the condensate takes place.

We now derive a relation between the diffusion time, $t_D$, and
the time of the condensation. From Eqs.(\ref{orig}) and
(\ref{solev}) we obtain
\begin{eqnarray*}
t_c\sim\frac{2\sigma_k^2}{D_0} \hspace{1.5cm} and \hspace{1.5cm}
t_D\sim\frac{k^2}{2D_0} \ ,
\end{eqnarray*}
where $\sigma_k=\frac{1}{2r_0}$, as follows from the Wigner
function
$N(\vec{k},\vec{r})=N_0exp\Bigl(-\frac{r^2}{2r_0^2}-\frac{k^2}
{2\sigma_k^2}\Bigr)$. Note that $r_0$ is the initial cross size of
the photon bunch.

Finally, we arrive at the desired relation
\begin{eqnarray}
\frac{t_D}{t_c}=k^2r_0^2 \ .
\end{eqnarray}
From here it is evident that for the condensation and evaporation
of photons, it is necessary that the following inequality $t_D\gg
t_c$ is satisfied. This is to be expected, as in realistic
astrophysical objects with a strong radiation, as well as in
laboratory experiments with a laser radiation, the following
condition $k^2r_0^2\gg 1$ is valid.

\section{SUMMARY AND DISCUSSIONS}

We have investigated a class of problems involving the interaction
of spectrally  broad and relativistically intense EM radiation
with a plasma in the case when the photon-photon interaction
dominates the photon-particle interactions. We have presented a
new concept of the establishment of equilibrium between the photon
and the wavepacket of EM field (the dense photon bunch). This is a
fully relativistic effect due to the strong EM radiation. We have
established the condition under which the variation of the plasma
density can be neglected in comparison with the variation of the
photon density. In such case the elementary excitations represent
the photonikos, for which we have derived the well known
Bogoliubov energy spectrum. We have studied the BEC and
evaporation of the photons from the Bose condensate in the case
when the density of plasma does not change. To this end, from the
Wigner-Moyal-Tsintsadze equation
\cite{ltsin98},\cite{ntsin98},\cite{men} we have derived a new
version of the Pauli kinetic equation for the photon gas. For the
case, when the wavevector and frequency of the photoniko is small
in comparison with the  wavevector and frequency of the photon, we
have derived the Fokker-Planck equation for photonikos. We have
presented a simple model, which exhibits the possibility of the
creation of Bose-Einstein condensate and evaporation of photons
from the condensate. We think that these processes can be
detectable in next generation experiments with appropriate
instrumentation. In fact, a number of experiments have been
carried out in which plasmas are irradiated by laser beams with
intensities up to $3\cdot 10^{20}W/cm^2$. At such intensities the
photon density is of the order of $n_\gamma\sim 10^{29}cm^{-3}.$
For the plasma densities up to $n_e\sim 10^{19}$, we should expect
that the Bose-Einstein condensate becomes observable. The theory
developed in this paper should also be the case for astrophysical
objects, such as a black hole. In this connection we speculate
that the recently observed radiation from the black hole may be
attributed to the evaporated photons from the Bose condensate.
That is initially all photons fall into Bose-Einstein condensate,
and then after a certain time, as discussed in this paper, some
photons undergo evaporation from the condensate. These processes
may also explain the observable variation in radiation intensity,
from being undetectable to one of the brightest sources on the
sky. Note that these sources can turn off for decades, and new
ones are always being found. In addition Universe is filled with a
gas of photonikos. This gas may play the decisive role in the
expansion of the Universe. Moreover, it may also be useful for
explaining certain processes in supernovae explosion. Finally, the
present theory may also find a valuable application in the future
space technology, as well as in nonlinear optics.

\begin{figure}
\caption{a) Absorption of the photoniko by the photon\\
b) Emission of the photoniko by the photon  }
\end{figure}

\end{document}